\documentclass[a4paper,fleqn]{cas-dc}

\usepackage[numbers,compress]{natbib}

\def\tsc#1{\csdef{#1}{\textsc{\lowercase{#1}}\xspace}}
\tsc{WGM}
\tsc{QE}
\tsc{EP}
\tsc{PMS}
\tsc{BEC}
\tsc{DE}

\begin{document}
\let\WriteBookmarks\relax
\def\floatpagepagefraction{1}
\def\textpagefraction{.001}
\shorttitle{}
\shortauthors{T.K Bhowmik et~al.}

\title[mode = title] {Al-dependent electronic and magnetic properties of YCrO$_{3}$ with magnetocaloric application: an ab-initio and Monte- Carlo approach}                      

\author[1] {Tushar Kanti Bhowmik}
\cormark[1]
\ead{tkb746@gmail.com}

\address[1] {Bose Institute, Department of Physics, 93/1, APC Road, Kolkata- 700009, India}

\author[1] {Tripurari Prasad Sinha}

\cortext [cor1] {Corresponding author}

\begin{abstract}
In this paper, a theoretical journey from electronic to magneto-caloric effect has been shown through the magnetic properties of aluminium induced yttrium chromate. The ground state electronic band structure and density of states have been studied using first principle calculations under GGA+U schemes. From the energy minimization, the ferromagnetic structure is more stable than the antiferromagnetic one. The interaction constant as well as the magnetic moment, have been determined from mean-field theory and DFT, respectively. The Monte-Carlo simulation under Metropolis algorithm has been employed to determine the critical temperature ($T_C$), which is nearly same as the experimental value. The temperature-dependent magnetization shows that these materials exhibit a paramagnetic to ferromagnetic phase transition at ~136 K, 130 K, 110 K, and 75 K respectively. The two inherent properties named the isothermal entropy change ($\Delta S_M$) and the adiabatic temperature change ($\Delta T_{ad}$) as a function of temperature for different applied magnetic fields have been determined to measure the magnetocaloric efficiency of these materials. The relative cooling power (RCP), which is calculated around $T_C$, changes from 4.7 J/Kg to 2.5 J/Kg with the decreasing Cr- concentration.
\end{abstract}

\begin{keywords}
Perovskite \sep DFT \sep Monte-Carlo Simulation \sep Critical Temperature \sep Magnetocaloric effect \sep RCP
\end{keywords}

\maketitle
\section{Introduction}

The ferromagnetic insulators  \cite{hena2008, Baidya2011, Kim2009, Yi2014} as well as semiconductors \cite{rogado2005,Chen2019,XIAO2018,Tan2019} within the group of perovskites have been drawing the tremendous interest in material research since last few decades due to their multi-functional properties such as magneto-capacitance \cite{Goto2004,Mori2005}, magneto-dielectric \cite{Singh2007,Maiti2013}, magneto-resistance \cite{KIM2020,Yamada2019}, magnetocaloric effect (MCE) \cite{Roy2016,Pakhira2017} etc. They have been applied in various fields such as spintronics \cite{Chang2019}, opto-electronics \cite{Fu2019}, piezoelectric \cite{You2017}. However, we have focused on the magnetocaloric material, which is used in the magnetic refrigeration for cooling at room temperature as well as the cryogenic application \cite{foldeaki1995,GIFFORD1970}. In the common refrigerator the cooling is done by compressing the vapour cycle \cite{Mascheroni2011}, which is very harmful for environment. But in magnetic refrigerator it is accomplished by application or removal of external magnetic field, which induces the magneto-caloric effect in these materials \cite{Mezaal2017}. The adiabatic temperature change ($\Delta T_{ad}$) and the isothermal magnetic entropy change ($\Delta S_m$), which are the intrinsic properties of the material, are the two important factors for the MCE. The magnetic refrigerator needs the magnetic materials, whose critical temperature ($T_C$) have within the range of requiring cooling temperature. So, we have been searching for the low cost magnetic perovskites, which have large MCE at transition temperature and provides the better degree of efficiency without affecting the environment.    

To know the underlying physics behind these strongly correlated materials, the density functional theory (DFT) \cite{Kohn1965} is beneficial and accurate until now. Using generalised gradient approximation (GGA) method \cite{Perdew1996} it can predict various properties of matter. It can be accurately described the d and f- electrons of strongly correlated systems by adding the Coulomb potential (U) with DFT (DFT+U) \cite{Anisimov1997}. But this theory can not count the thermal fluctuations present in the system. But, it is necessary to discuss the temperature-dependent behaviour of this type of material. In the critical region, where the phase transition occurs, the DFT can not explain the physics behind it.  There are several types of theoretical models such as mean-field theory \cite{Toulouse1981}, 3D- Heisenberg model \cite{Holm1993}, 3D Ising model \cite{Talapov1996}, etc.,  to describe this behaviour accurately. These models are enabled to explain the mechanism near phase transition and help us to find the appropriate materials for MCE. Usually, the Monte-Carlo simulation technique has been used to describe these models \cite{Binder1993}. 

In this context, we have targeted to reach the liquid nitrogen temperature (77 K), so that we can easily replace liquid nitrogen as a frigging element, which is very costly and uneasy to handle. So, we use the yttrium chromate (YCrO$_3$)(T$_C$ = 140 K) \cite{JARA2018} as a starting material and then we doped a nonmagnetic aluminium (Al) to Cr site to decrease the critical temperature. With 50\% doping of Al \cite{DURAN2018}, we have achieved that the T$_C$ is 73 K. At first, we have used DFT to know the Fermi energy, the density of states (DOS), bandgap energy. We have studied the band structure and DOS of 10\% and 30\% doping of Al with full potential linear augmented plane wave (FPLAPW) method \cite{ARAY2002}. From DFT, we can easily predict the ground state magnetic configuration of these materials. After that, we have used the Ising model to determine the transition temperatures (T$_C$), magnetic susceptibility ($\chi$), and hysteresis loop with the help of Metropolis Monte- Carlo algorithm. At last, we have calculated the adiabatic temperature change ($\Delta T_{ad}$) as well as the relative cooling power (RCP) to determine the magnetocaloric efficiency of these systems.                 

\begin{figure*}
	\centering
		\includegraphics[scale=.40]{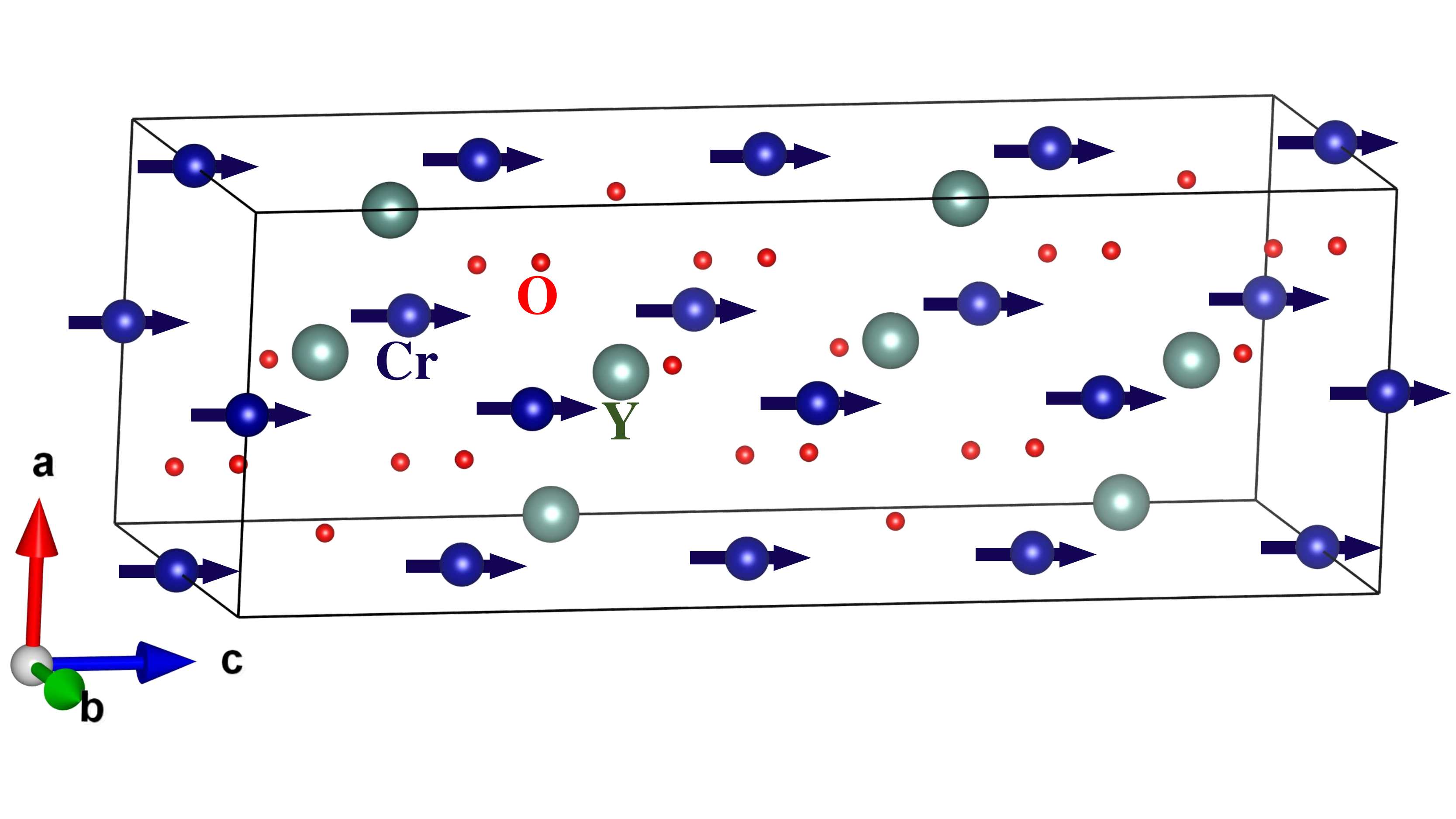}
	\caption{Crystal structure of YCrO${_3}$, used for theoretical calculations and the blue arrows signify the spin direction of Cr-atom with FM configuration.}
	\label{Figure:1}
\end{figure*}

\section{Computational Details}
\subsection{Ab-initio Calculations}
Density functional theory (DFT) is implemented to study the electronic structure and magnetic properties of Al-doped YCO$_3$. The full- potential linearized augmented plane wave (FPLAPW) method have been used to solve the Kohn-Sham equations as implemented in WIEN2K package \cite{BLAHA1990,Blaha2020}. The general gradient approximation (GGA) with Coulomb repulsion U (GGA+U) method has been employed to optimize the crystal structures as well as determining the electronic structure and the magnetic moment calculations. We have used different supercell of 40 atoms with P1 spacegroup and replaced the Cr atom with Al atom appropriately to construct the YAl$_{0.1}$Cr$_{0.9}$O$_3$ and YAl$_{0.3}$Cr$_{0.7}$O$_3$ structures. The muffin- tin raddi ($R_{mt}$) for Y, Al, Cr and O atoms are 2.26 {\AA}, 1.89 {\AA}, 1.78 {\AA} and 1.69 {\AA} respectively. The cut-off energy ($R_{mt}$ $\times$ $K_{max}$) from core states to valance states has been set to 6 eV. 500 k- points have been used in the whole Brillion zone for self- consistent convergence. The Hubbard parameter $U_{eff}$ = $U$ - $J$ = 4 eV have been used for Cr- 3d states. The energy convergence criteria are set to $10^{-4}$ Ry, whereas the charge convergence is fixed to $10^{-3}$ e for self- consistent optimize calculations of Al-doped YCrO$_3$.

The electronic structure of Al-doped YCrO$_3$ has been calculated using the GGA+U method. It is reported that YCrO$_3$ crystallizes with the orthorhombic Pnma space group. In the case of Al-doped YCrO$_3$, the supercell has been generated in WIEN2k struct file and replaced the Chromium with Aluminium appropriately to maintain the doping concentration. The spin direction of the magnetic Cr-ions are set to the upward direction for FM calculations. In order to see the magnetic structure, we have tried with AFM spin configuration between Cr-ions also. According to the energy value of these FM and AFM configurations, we have seen the FM-spin obtained the minimal energy value for all cases. The detailed magnetic properties have been described in the following sections. The calculated structure has been optimized according to WIEN2k spin-polarized algorithm. After that, the density of states and band structure have been calculated.

\subsection{Monte-Carlo Simulations}

Yttrium Chromate (YCO) is a single perovskite ABO$_3$ like structure where A site cation Y$^{3+}$ is non magnetic but B site cation Cr$^{3+}$ is a magnetic one. So we have considered the magnetic interaction between Cr$^{3+}$- Cr$^{3+}$ ions only. The other three samples we have doped randomly the non magnetic Al$^{3+}$ ions in different ratios in Cr-site. The interaction constant between Cr$^{3+}$ and Al$^{3+}$ is taken to zero due to the non magnetic Al$^{3+}$ ion. So we have considered only Cr$^{3+}$ - Cr$^{3+}$ nearest neighbour (nn) and next nearest neighbour (nnn) interaction here. We have taken Ising model Hamiltonian with nn and nnn for the simulation.

\begin{equation}
    H = -J_1\sum_{<i,j>}{s_is_j} -J_2\sum_{<<i,k>>}{s_is_k} -h\sum_{i}{s_i}
\end{equation}
Where $J_1$ and $J_2$ are the nn and nnn interaction constants and $<i, j>$ and $<<i, k>>$ are nn and nnn sites respectively. 
From previous experimental study of YCO, we have seen that it crystallizes with orthorohmbic Pnma space group where the Cr$^{3+}$ ion forms CrO$_6$ octahedra with all six nearest O$^{2-}$ ions. The antiferromagnetic ordering temperature is T$_N$ = 140 K but bellow ordering temperature hysteresis curve have been shown that there was some week ferromagnetism present in this material. In our magnetic monte carlo simulation study, we have considered only magnetic Cr$^{3+}$ ion, which positioned at the corner of the  cubic crystal structure and nn interaction between two Cr- ion have chosen to antiferromagnetic where as nnn interaction is considered to be a ferromagnetic. We have used the mean field approximation \cite{Van1941} to determine the interaction constants $J_1$ (nn) and $J_2$ (nnn). 
\begin{equation}
    T_N = \frac{2}{3k_B}ZS(S+1)J_{eff}
\end{equation}
The value of the transition temperatures (T$_N$ = 140 K) are taken from the previous experimental study and Z are the co-ordination number (Here Z = 6 and S = 3/2). The values of $J_{eff}$, which have been calculated from equation (2), are 9.33 K, 9.85 K, 12.78 K and 13.9 K for YCrO$_3$, YAl$_{0.1}$Cr$_{0.9}$O$_3$, YAl$_{0.3}$Cr$_{0.7}$O$_3$ and YAl$_{0.5}$Cr$_{0.5}$O$_3$ respectively.


\begin{figure*}
	\centering
		\includegraphics[scale=.67]{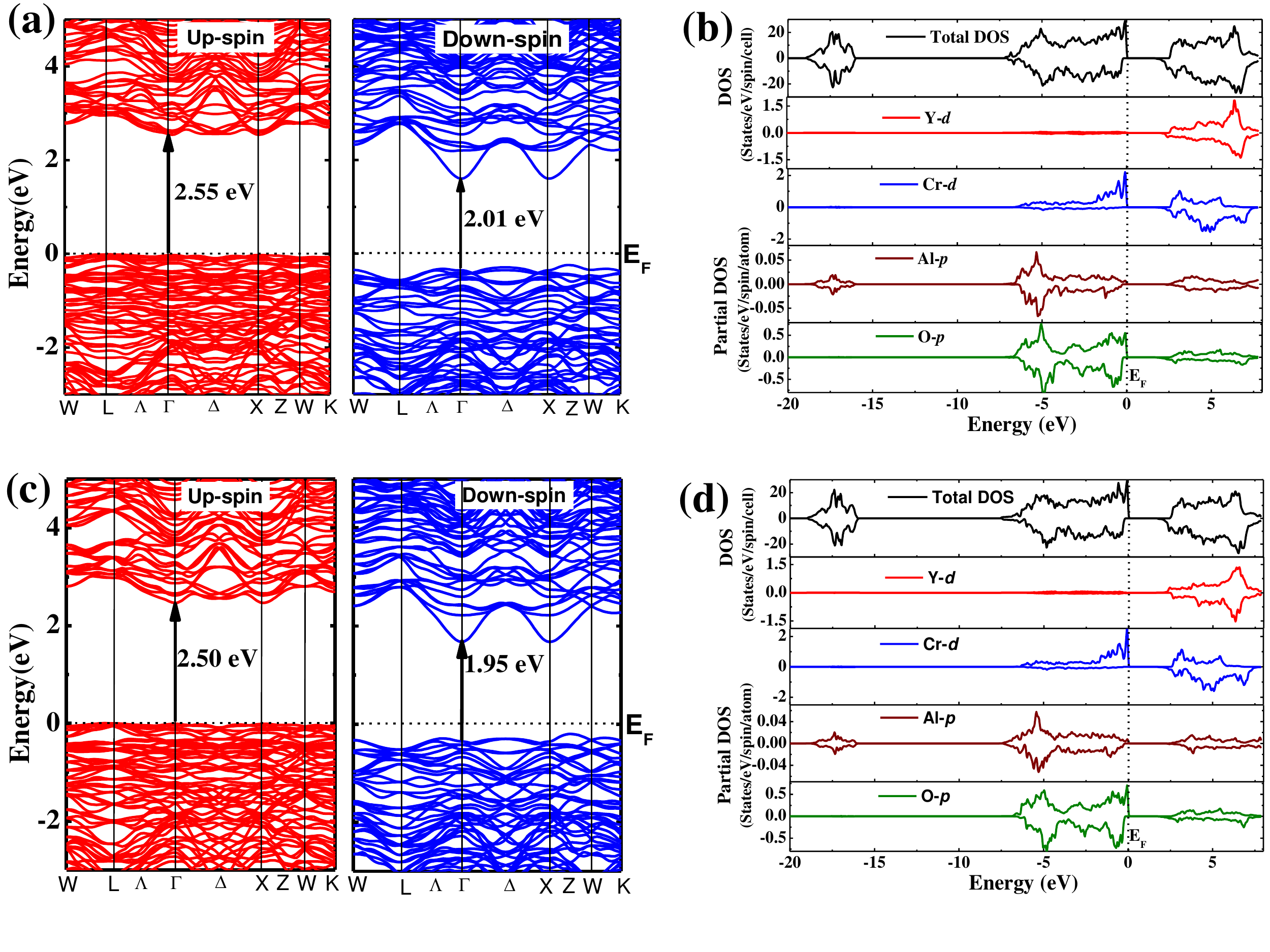}
	\caption{Band structure from GGA+U method for up spin (red) and down spin (blue) configuration for (a) YCr$_{0.9}$Al$_{0.1}$O$_3$ (c) YCr$_{0.7}$Al$_{0.3}$O$_3$. Density of states (DOS) and Partial DOS for different atoms for (b) YCr$_{0.9}$Al$_{0.1}$O$_3$ (d) YCr$_{0.7}$Al$_{0.3}$O$_3$}.
	\label{Figure:2}
\end{figure*}

\begin{figure*}
	\centering
		\includegraphics[scale=.38]{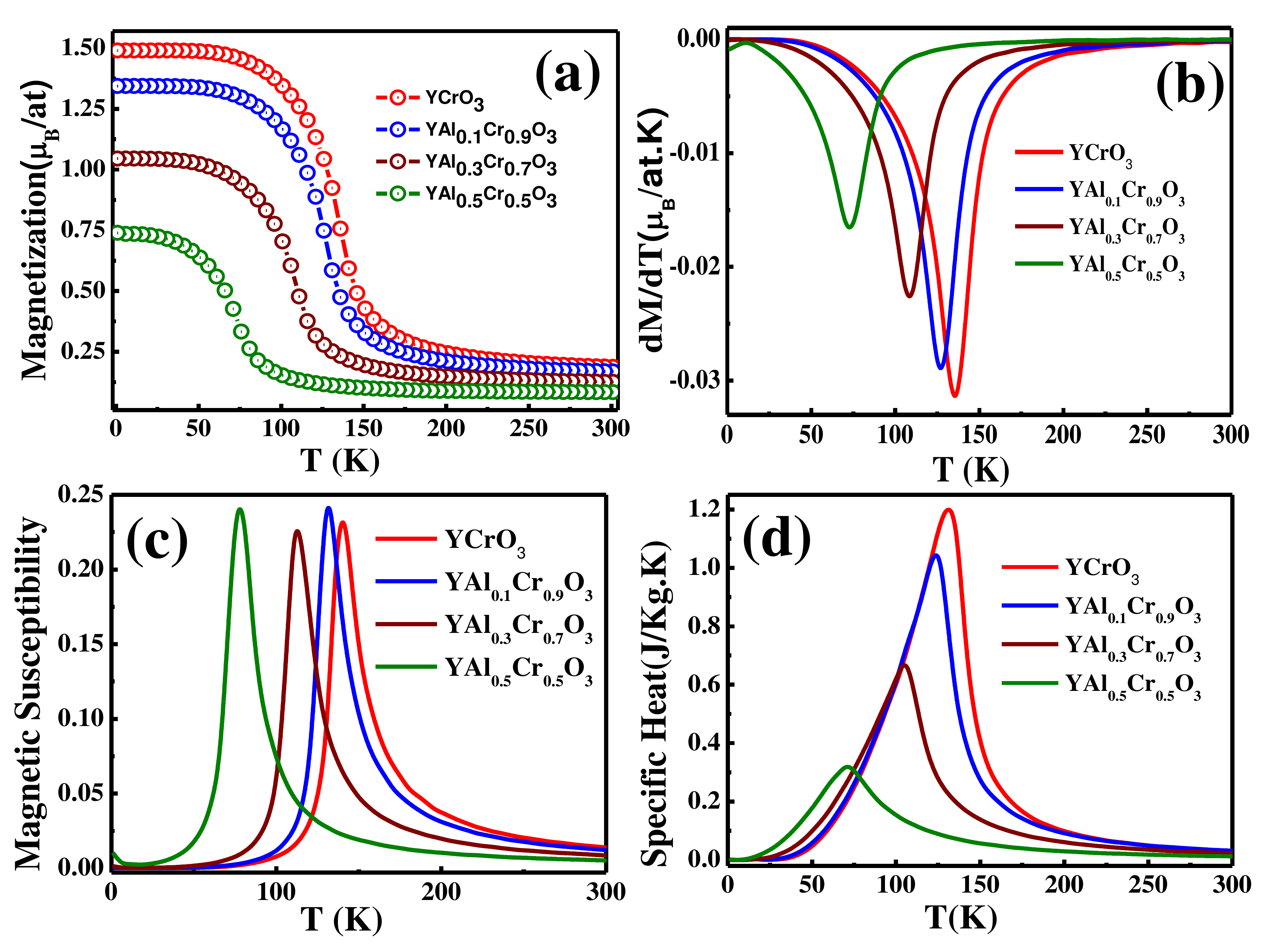}
	\caption{(a) Temperature dependent magnetization for different Al doping concentration. (b) 1st derivative of M vs T curves for all materials. (c) Magnetic susceptibility vs temperature curves and (d) Specific heat with respect to temperature for YCr$_{1-x}$Al$_x$O$_3$.}
	\label{Figure:3}
\end{figure*}

\begin{figure*}
	\centering
		\includegraphics[scale=.38]{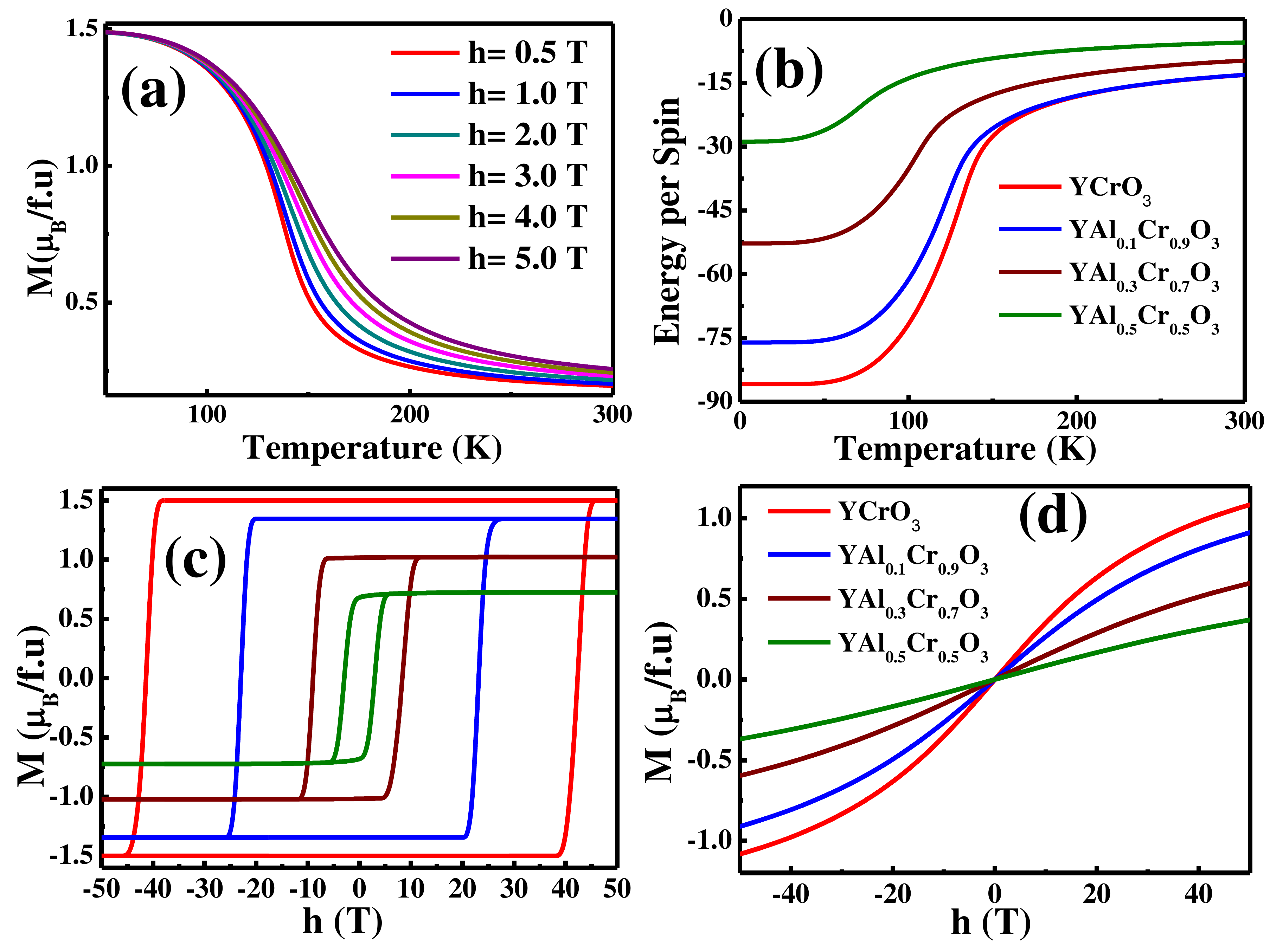}
	\caption{ (a) Magnetization versus temperature at different external field for YCO$_3$. (b) Temperature dependency of energy per spin of differnt samples. (c) M-H loop at 20 K. (d) M-H curve at room temperature for different materials.}
	\label{Figure:4}
\end{figure*}

We have done Monte-Carlo simulation (MCS) using the Metropolis algorithm to determine the magnetic properties of YAl$_{x}$Cr$_{1-x}$O$_3$ (x = 0, 0.1, 0.3, 0.5). In the MCS process, the periodic boundary condition and standard sampling method have been applied over the whole lattice of size L = 30 in all three Cartesian directions. Single flip metropolis algorithm has been used to solve the above Ising Hamiltonian, described at equ (1). If the change in energy (dE) is less than zero, or the transition probability [$exp(-dE/k_BT)$] is greater than a random value, the flip is accepted otherwise rejected. 10$^6$ MCS steps have been used to reach the equilibrium of lattices and next 10$^7$ steps to average the magnetization and other observables. The physical quantities, which have been measured using MCS, are described as follows. The internal energy per site is,
\begin{equation}
    E = \frac{1}{N}<H>
\end{equation}
Where $N = L\times L\times L$ is the total number of lattice points.
The magnetisation of Cr3+ ions in these materials is, 
\begin{equation}
    M = <\frac{1}{N}\sum_i S_i>
\end{equation}
The magnetic susceptibility is given by,
\begin{equation}
    \chi = \frac{<M^2>-<M>^2}{k_BT}
\end{equation} 										
The magnetic specific heat is calculated by,
\begin{equation}
    C_m = \frac{<E^2>-<E>^2}{k_BT^2}
\end{equation}  					
The magnetic entropy is given by
\begin{equation}
    S_m = \int_0^T \frac{C_m}{T} dT
\end{equation}
The change in magnetic entropy is described as
\begin{equation}
    \Delta S_m = \int_{0}^{h_{max}} (\frac{\partial M}{\partial T}) dh
\end{equation}
Where $h_{max}$ is the maximum applied field and $(\frac{\partial M}{\partial T})_h$ is the thermal magnetisation for a fixed external magnetic field $h$. The adiabatic temperature change is given by
\begin{equation}
    \Delta T_{ad} = -T \frac{\Delta S_m}{C_m}
\end{equation}
The magnetocaloric effect is measured by a parameter called Relative Cooling Power (RCP), which is calculated from the magnetic entropy change vs temperature curves. The formula for determining of RCP is given by,
\begin{equation}
    RCP = \int_{T_1}^{T_2} \Delta S_m (T) dT
\end{equation}	 									
Where, $T_1$ and $T_2$ are the cold and hot temperature corresponding the both ends of half maximum value of $\Delta S_m$ vs $T$ curves.

\section{Results and Discussions}
\subsection{Electronic Properties}

The electronic density of states (DOS) and band structures have been calculated after optimizing the crystal structure of Al-doped YCrO$_3$. The band structures and DOS of YAl$_{1-x}$Cr$_x$O$_3$ (x = 0.1 and 0.3) have been shown in figure 2. Spin-polarized GGA+U method has been applied to determine the energy band of these systems. Figure 2(a) and 2(c) represent the band structure, which has been calculated along with the high symmetry points in Brillouin zone for the majority (UP) and minority (DOWN) spin direction, of YAl$_{0.1}$Cr$_{0.9}$O$_3$ and YAl$_{0.3}$Cr$_{0.7}$O$_3$ respectively. The dotted line corresponds the Fermi energy ($E_F$), which separates the valance band and conduction band of Al-doped YCrO$_3$. The majority spin band gaps at $\Gamma$ points (direct bandgap, $\Gamma$ $\rightarrow$  $\Gamma$) are 2.55 eV and 2.5 eV for 10\%  and 30\% Al-doped samples. The UP spin direct bandgap for YCrO$_3$ is 2.61 eV, which means the bandgap is decreasing with increasing Al concentration. The experimental band gap of YCO$_3$ is reported to 1.86 and 2.86 eV \cite{Saxena2018}, which is nearly same as our theoretical value 1.90 eV and 2.61 eV. The minority spins have shown the indirect bandgap ($Z \rightarrow \Gamma$) which are 1.9 eV, 1.95 eV, 1.951 eV and 2.08 eV respectively for x= 0.0, 0.1, 0.3, 0.5 doping concentrations.

To study more intensely about the energy states below and above the Fermi level, we have calculated the partial DOS (PDOS) of Y-d, Cr-d, Al-p and O-p states. The total DOS and PDOS have been described in figure 2(b), and 2(d) for YAl$_{0.1}$Cr${_0.9}$O$_3$ and YAl$_{0.3}$Cr$_{0.7}$O$_3$, respectively. The conduction band mainly consists of Y-d states as well as Cr-d states. The spin up and down channel, the minima of the conduction band at $\Gamma$ and X point originated due to the Cr-d states. O-p states are responsible for the valance band maxima in the spin-down channel at Z point. The Al-p, O-s states are responsible for lower band bellow -8 eV. The asymmetry in the up and down spin channel for Cr-d states confirms that it is ferromagnetic in nature.

\subsection{Magnetic Properties}

The ground state magnetic properties of Al-doped YCrO$_3$ compounds have been studied using the spin polarization GGA-PBE+U method. These systems are optimized with two different magnetic spin systems of the magnetic Cr-atom. The interaction between two nearest neighbours Cr-atom is considered to be a ferromagnetic (FM) and antiferromagnetic (AFM), respectively. It is seen that the energy eigenvalues of the FM configuration are less than the AFM for all cases. So, the ground state spin structure of these compounds are ferromagnetic. The magnetic moments per Cr atom for FM configuration are 1.5 $\mu_B$, 1.31 $\mu_B$, 1.13 $\mu_B$, 0.75 $\mu_B$ for YAl$_{x}$Cr$_{1-x}$O$_3$, (where x = 0.0, 0.1, 0.3, 0.5) respectively.

Figure 3(a) represents the magnetization per Cr-atom of YAl$_{x}$Cr$_{1-x}$O$_3$, (where x = 0.0, 0.1, 0.3, 0.5) as a function of temperature without any external magnetic field. The paramagnetic transition temperature ($T_C$) decreases with increasing Al doping concentrations. To determine the exact transition temperature, we have differentiated the magnetization with respect to temperature and plotted the obtained $\frac{dM}{dT}$ curve with temperature (T) in figure 3(b). The minimum point of this curve indicates the magnetic ordering temperatures at 136 K, 126 K, 110 K and 73 K for YAl$_{x}$Cr$_{1-x}$O$_3$ compound with x = 0.0, 0.1, 0.3 and 0.5 respectively. The obtained values of transition temperature are comparable to the experimental results. The saturation magnetization ($M_S$) decreases linearly with this equation, $M_S = M_{S_0}(1-x)$, where $M_{S_{0}}$ is the saturation magnetization of starting material YCO and x is the different Al concentrations. The temperature-dependent magnetic susceptibility and specific heat are shown in figure 3(c) and 3(d) for different Al doping concentrations. These two observables are calculated from the magnetic and energy fluctuation, respectively, with the help of equation (5) and (6). The peaks in these two figures signify the transition temperature of these materials, which is as same as calculated earlier.

Figure 4(a) shows the temperature dependence of magnetization in the presence of the different applied external magnetic field, where the steepness of the curve depends on the applied magnetic field. It is seen that the saturation magnetization is the same for all applied field, but the transition temperature shifts towards the higher values for increasing of the applied field. The increase in $T_C$ indicates that the external magnetic field influences the magnetic interaction between two nearest neighbour magnetic ion. The alignment of all spins of Cr$^{3+}$ ion becomes uni-direction, i.e. towards the external field direction before the transition temperature upon cooling of this material. The ground state magnetic energy per spin is shown in figure 4(b). The ground state magnetic energy is calculated from mean-field theory using the relation $E_0 = -\frac{1}{2}Z_{eff}J_{eff}S_{eff}^2$ \cite{Diep2014}. The values of ground-state free energy are -86.94 meV, -76.71 meV, -53.01 meV and -28.84 meV for x= 0, 0.1, 0.3 and 0.5 respectively. From our simulation, we have seen the values of free energies are -85.89 meV, -76.08 meV, -52.78 meV and -28.82 meV respectively, which is very similar to the calculated values. The big change of slope in the curves indicates the phase transitions temperature of these materials. Figure 4(c) and (d) show how the magnetization changes with the applied magnetic field from 50 T to -50 T at 20 K and 200 K respectively. From 20 K hysteresis loop, we see that the remnant magnetization as well as the coercive field both decreases with increasing Al concentrations. The Al doping also causes the reduction of the saturation magnetization ($M_S$) bellow the transition temperature. Below the transition temperature, it shows the wide gap between negative and positive values of the applied field, which confirms the ferromagnetic nature of these materials. Above the transition temperature, they have not shown any remnant magnetization as well as the coercive field. This type of behaviour supports the paramagnetic nature of these materials.      
      
\begin{figure*}
    \centering
        \includegraphics[scale=.36]{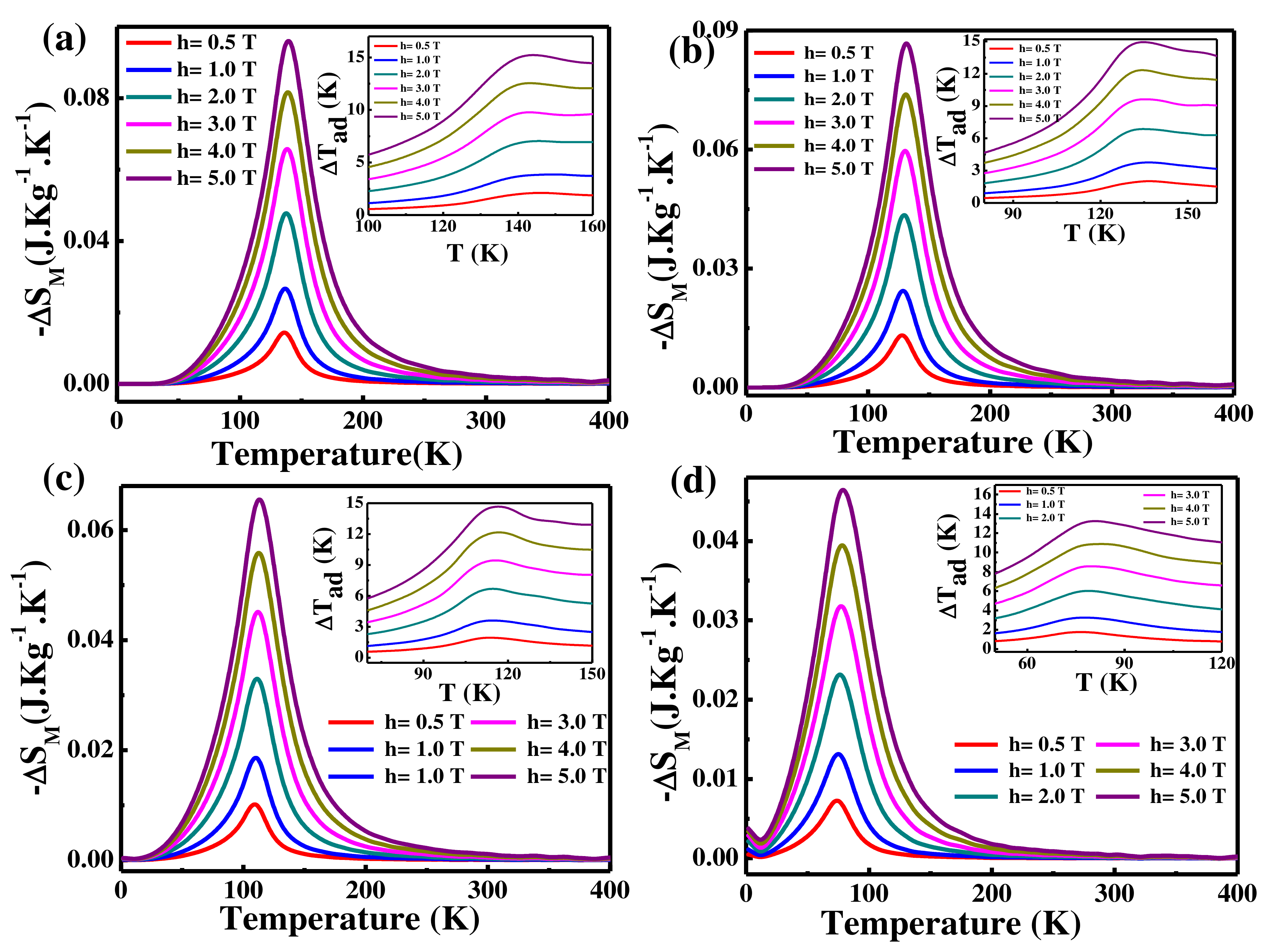}
    \caption{ Negative magnetic entropy changes with respect to temperature and the inset shows the temperature dependence of adiabatic temperature change for YCr$_{1-x}$Al$_x$O$_3$ with different external magnetic field (a) x = 0.0, (b) x = 0.1, (c) x = 0.3 (d) x = 0.5}
    \label{Figure: 5}
\end{figure*}

\begin{figure}
    \centering
    \includegraphics[width=6.5cm,height=8.5cm]{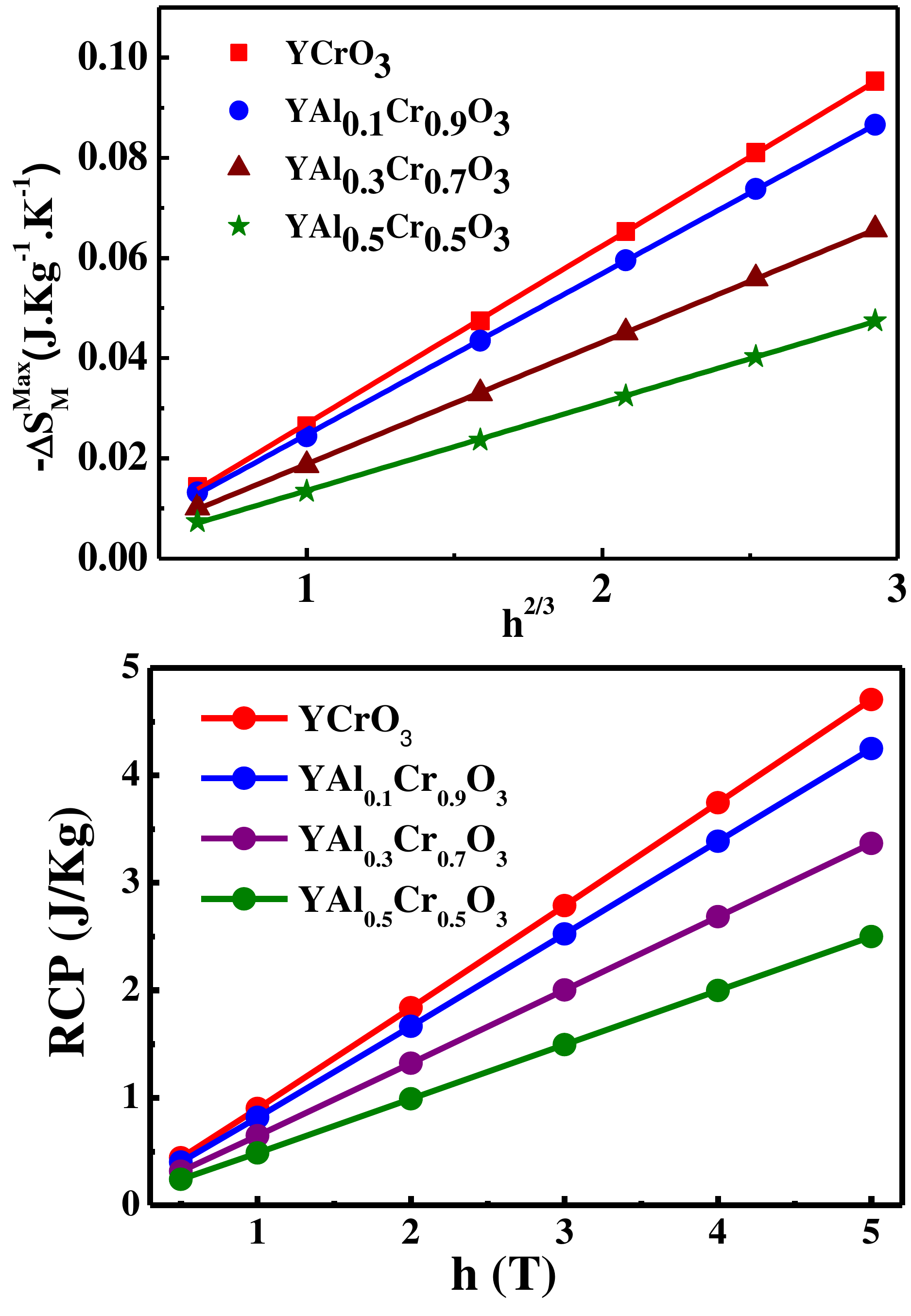}
    \caption{(a) Maximum entropy change with respect to the two third of the external magnetic field. (b) RCP versus external magnetic field for different materials.}
    \label{Figure: 6}
\end{figure}

\subsection{Magnetocaloric Effect}

The magnetic entropy change (-$\Delta S_m$) with respect to temperature for different applied magnetic field has been shown in figure 5(a-d) for YAl$_{x}$Cr$_{1-x}$O$_3$, x = 0.0, 0.1, 0.3 and 0.5. The symmetric peak of these curves indicates the second-order phase transition. It shows the positive value over the all temperature range around the wide range of the transition temperature, which is very useful for magnetic refrigeration around 136 K, 126 K, 110 K and 73 K. For an active regenerative magnetic refrigerator, we need a series of different $T_C$ materials. So, we have doped the non-magnetic Al in different composition in the Cr site to reduce transition temperature gradually towards the lowering temperature. The magnitude, as well as the full-width at the half maximum (FWHM) of the curve, is gradually increasing with the applied field. The value of FWHM ($\Delta T$) is a very useful parameter to determine the operating temperature range of the magnetic refrigeration. In our case the value of FWHM of 5 tesla external field is 49.39 K, 49.07 K, 51.23 K and 52.66 K for YAl$_{x}$Cr$_{1-x}$O$_3$, x = 0.0, 0.1, 0.3 and 0.5 respectively. The maximum value of -$\Delta S_m $ as a function of two-third power of applied field ($h^{2/3}$) increases linearly, which indicates again that it is the second-order phase transition, shown in figure 6(a). 

The adiabatic temperature change ($\Delta T$) in the presence of the magnetic field has been calculated from magnetic entropy difference and magnetic specific heat. The inset of figure 5(a-d) is described this adiabatic temperature change, which is maximum at the transition temperature. The values of $\Delta T_{max}$ which is proportional to the applied magnetic field are 15.3 K, 15 K, 14.6 K and 13.3 K of 5 T magnetic field for YAl$_{x}$Cr$_{1-x}$O$_3$, x = 0.0, 0.1, 0.3 and 0.5 respectively. This is a very useful parameter which describes the amount of cooling performance of the material. The value tells us that it is very effective as magneto-caloric refrigerator applications.

Finally, we have measured the relative cooling power (RCP) of these set of materials as a function of the applied magnetic field. This parameter plays a vital role to characterise the magneto-caloric material as well as to determine the efficiency of magnetic cooling. The RCP value, which is shown in figure 6(b), increases linearly with increasing applied field. The RCP decreases with increasing Al concentrations.

\section{Conclusions}

In this current paper, we have studied electronic and magnetic properties of aluminium doped yttrium chromate using first principle calculations after that we have re-investigated the magnetic properties and studied the magneto-caloric effect of these systems using Monte- Carlo simulations. We have seen that the bandgap of these materials are indirect due to the valance band maxima (Z- point) and conduction band minima $\Gamma$ point are different point of symmetry in the Brillouin boundary. The calculated magnetic moment of Cr atom from ab-initio study and Monte-Carlo simulations at ground state is the same one. The ferromagnetic transition temperature (T$_C$) of these materials decreases with increasing Al- concentrations and simulated T$_C$ is as same as the experimental value. The magnetic entropy change ($\Delta S_m$) and adiabatic temperature change ($\Delta T$) have been calculated to study the magneto-caloric effect of these materials. The RCP value increases linearly with the application of the higher external magnetic field. We propose that we can easily reach into the liquid nitrogen temperature (78 K) using these series of Al-doped YCrO$_3$ as the magnetic elements of a magnetic refrigerator.

\section{Acknowledgement}

T.K Bhowmik would like to thank Department of Science and Technology (DST), Government of India for providing the financial support in the form of DST-INSPIRE fellowship (IF160418). 

\bibliographystyle{model1-num-names.bst}
\bibliography{YACO}

\end{document}